\documentclass[sn-nature,xcolor={usenames}]{sn-jnl}
\usepackage{times}

\usepackage{xr}  

\usepackage{geometry} 

\usepackage[numbers]{natbib}
\usepackage{multicol}
\usepackage{graphicx}%
\usepackage{multirow}%
\usepackage{amsmath,amssymb,amsfonts}%
\usepackage{amsthm}%
\usepackage{mathrsfs}%
\usepackage[title]{appendix}%
\usepackage{xcolor}%
\usepackage{textcomp}%
\usepackage{manyfoot}%
\usepackage{algorithm}%
\usepackage{algorithmicx}%
\usepackage{algpseudocode}%
\usepackage{listings}%

\usepackage{subcaption}
\usepackage{siunitx}

\definecolor{notes}{RGB}{200,0,0}
\definecolor{newChanges}{RGB}{0,0,0}
\definecolor{newerChanges}{RGB}{0,0,0}
\definecolor{question}{RGB}{255,0,0}
\definecolor{ToDo}{RGB}{153, 51, 0}

\definecolor{dark-red}{rgb}{0.4,0.15,0.15}
\definecolor{dark-blue}{rgb}{0.15,0.15,0.8}
\definecolor{medium-blue}{rgb}{0,0,0.5}

\newcommand{\newChanges}[1]{{\color{newChanges}{#1}}}
\newcommand{\newerChanges}[1]{{\color{newerChanges}{#1}}}

\usepackage{tikz}
\newcommand\copyrighttext{%
  \footnotesize  \textcopyright Accepted at Nature Portfolio Journal (NPJ) Complexity}
\newcommand\copyrightnotice{%
\begin{tikzpicture} [remember picture,overlay]
 \node[anchor=south,yshift=10pt] at (current page.south)
    {\copyrighttext};
\end{tikzpicture}%
}

\usepackage{soul}  

\begin{document}

\title{Messengers: Breaking Echo Chambers in Collective Opinion Dynamics with Homophily}

\author*[1,2,3]{\fnm{Mohsen} \sur{Raoufi}}\email{mohsenraoufi@icloud.com}

\author[1,4]{\fnm{Heiko} \sur{Hamann}}\email{heiko.hamann@uni-konstanz.de}

\author[1,2]{\fnm{Pawel} \sur{Romanczuk}}\email{pawel.romanczuk@hu-berlin.de}

\affil[1]{\orgdiv{Science of Intelligence}, \orgname{Research Cluster of Excellence}, \orgaddress{\street{Marchstr. 23}, \city{Berlin}, \postcode{10587},  \country{Germany}}}

\affil[2]{\orgdiv{Department of Electrical Engineering and Computer Science}, \orgname{Technical University of Berlin}, \orgaddress{\street{Marchstr. 23}, \city{Berlin}, \postcode{10587},  \country{Germany}}}

\affil[3]{\orgdiv{Department of Biology}, \orgname{Humboldt University of Berlin}, \orgaddress{\street{Philippstr. 13}, \city{Berlin}, \postcode{10115}, \country{Germany}}}

\affil[4]{\orgdiv{Department of Computer and Information Science}, \orgname{University of Konstanz}, \orgaddress{\street{Box 188}, \city{Konstanz}, \postcode{78457},  \country{Germany}}}

\newcommand{\supplementaryurl}{https://doi.org/10.6084/m9.figshare.25982572.v1} 

\newcommand{\homophilyVideoUrl}{https://doi.org/10.6084/m9.figshare.26217539.v1}

\abstract{
Collective estimation is a variant of collective decision-making where agents reach consensus on a continuous quantity through social interactions.
Achieving precise consensus is complex due to the co-evolution of opinions and the interaction network.
While homophilic networks may facilitate estimation in well-connected systems,  disproportionate interactions with like-minded neighbors lead to the emergence of echo chambers and prevent consensus. Our agent-based simulations confirm that, besides limited exposure to attitude-challenging opinions, seeking reaffirming information entrap agents in echo chambers.
To overcome this, agents can adopt a stubborn state (Messengers) that carry data and connect clusters by physically transporting their opinion. We propose a generic approach based on a Dichotomous Markov Process, which governs probabilistic switching between behavioral states and generates diverse collective behaviors. We study a continuum between task specialization (no switching), to generalization (slow or rapid switching). Messengers help the collective escape local minima, break echo chambers, and promote consensus.
}


\maketitle

\section{Introduction}
\copyrightnotice
Collective behaviors exhibited by animal collectives or groups of interacting artificial agents are fascinating examples of self-organization. Through these behaviors, groups can solve problems collectively, that cannot be solved by any individual alone. This phenomenon is often referred to as collective or swarm intelligence \cite{bonabeau1999swarm,krause2010swarm,leonard2022collective}. While significant progress has been made over the past decades in understanding the fundamentals of collective intelligence, many questions remain open regarding actual mechanisms underlying collectively intelligent behavior, in particular in spatially embedded systems lacking the capability for global information exchange between individual agents. 

Among the many different manifestations of collective intelligence, the wisdom of crowds effect~\cite{galton_vox_1907, surowiecki2005wisdom, budescu2015identifying} stands out as a great candidate for studying collective computational intelligence~\cite{leonard2022collective}. While the core idea--that the average of many imperfect estimations can be remarkably close to the true value--seems straightforward, achieving precise estimations depends on meeting certain criteria~\cite{surowiecki2005wisdom, golub2010naive, prelec2017solution, da2020harnessing, winklmayr2020wisdom}. Typically, it assumes global knowledge by individuals, which is often not achievable in fully decentralized settings. However, distributed consensus models, particularly DeGroot-like models~\cite{degroot1974reaching}, implement the wisdom of crowds effect without a centralization assumption~\cite{olfati2007consensus, golub2010naive}. These models have also been used to model opinion dynamics, by providing a mechanism for how opinions spread across the interaction network~\cite{xia2011opinion, lorenz2018opinion, bizyaeva2022nonlinear, olsson2024analogies}. \newerChanges{Such models find applications in engineering domains like distributed sensing, collective robotics, and machine learning~\cite{xiao2005scheme, leonard2007collective, hamann2018opinion, ICRA2023Raoufi}, as well as provide insights into social systems for economics, politics, and collective behavior~\cite{zha2020opinion, lorenz2018opinion, golub2010naive, maia2021adaptive, castellano2009statistical}. In this work, we study a specific application of opinion dynamics that leads to collective estimation~\cite{ICRA2023Raoufi}, where agents integrate distributed information to arrive at accurate estimation. 
\par \noindent
To better understand the underlying interaction rules that drive the group-level behavior, many studies rely on agent-based models. These models capture how simple, local rules, when applied at the level of each agent, can lead to emergent consensus or disagreement at the group level. Common examples for opinion dynamics include the} Voter models~\cite{liggett1999stochastic}, bounded confidence~\cite{hegselmann2002opinion}, and gossip methods~\cite{sirocchi2022topological}. \newerChanges{Beyond the update rules themselves, the structure of interactions plays a critical role in shaping the outcome of opinion dynamics.} The underlying interaction network is an important determinant of collective behavior~\cite{sirocchi2022topological, galesic2023beyond}. \newerChanges{In particular, as the network becomes sparser, there exists a sparsity threshold above which the network fragments, making consensus unattainable. For example, in static networks, connectivity governs a tradeoff between the speed and accuracy of consensus formation: stronger connectivity leads to faster convergence but can reduce decision accuracy due to premature agreement~\cite{ICRA2023Raoufi, winklmayr2020wisdom}. This tradeoff becomes even more pronounced in spatially embedded systems, where agents must physically spread out to access diverse information, but increased dispersion risks breaking the network apart.}

\par
Contrary to the assumption of static networks in early models of opinion dynamics, many real-world systems exhibit dynamic networks, as agents tend to constantly rearrange their connections with their peers, making the collective behavior more complex~\cite{fisher2021using}. \newerChanges{These adaptive interactions introduce additional complexity into collective behavior, as the network structure evolves along with other system states. A prominent mechanism driving such rewiring is homophily: the disproportional tendency of agents to establish links with like-minded neighbors~\cite{mcpherson2001birds,karimi2018homophily,lee2019homophily}, which in turn reshapes information flow in the network.}
\par \noindent
One notable consequence of homophily is the emergence of `echo chambers': clusters of agents that are internally homogeneous in their opinions and disconnected from other clusters with different opinions. Echo chambers reinforce only the locally dominating perspective, foster confirmation biases, and prevent exposure to attitude-challenging information~\cite{tornberg2018echo, mocanu2015collective, bakshy2015exposure}. \newerChanges{These structures disrupt the flow of diverse information across the network and can lead to collective outcomes beyond simple fragmentation.} These effects have been linked to phenomena that are considered threats to public discourse and human society~\cite{del2016spreading,bak2021stewardship,muller2022echo}. For example, in the spread of misinformation \cite{bakshy2015exposure, lovato2024diverse}, local information homogeneity acts as a driver to the diffusion of misinformation \cite{del2016spreading, tornberg2018echo, stein2023network}. Other consequences include the formation of filter bubbles and network segregation \cite{maia2021adaptive, stein2023network}. 
\par \noindent
Despite their adverse effects at the collective level, echo chambers function for a purpose on the individual level: comforting agents with reaffirmation and protecting them from disagreement \cite{tornberg2018echo}. Although the actual adversarial outcome of echo chambers in collective dynamics is arguable \cite{levy2019echo}, \newerChanges{their structural role is clear. In this paper, where reaching consensus on distributed information is a key objective, echo chambers are barriers that inhibit information exchange and reduce the precision of collective estimation. 
In our collective estimation scenario, the presence of a defined objective allows us to examine the effects of homophily and echo chambers in measurable terms and to derive design principles for mitigation in artificial systems.
}

\par
In this work, we explicitly consider scenario of spatially embedded opinion dynamics, as encountered in real-world groups of mobile agents moving through physical space~\cite{raoufi2021speed,ICRA2023Raoufi}. 
\newerChanges{While many mechanistic models have been developed to describe the co-evolution of homophilic networks and opinion dynamics, leading to the emergence of echo chambers~\cite{holme2006nonequilibrium, bullock2023agent, zanette2006opinion, bullock2024spatial}, they typically focus on the evolution of opinions through social interactions, without considering where those opinions originate. In most of these models, agents' initial opinions are drawn from a random pool of discrete options.}
Therefore, by design, these models do not take into account the influence of external sources of information on the evolution of opinions, e.g., information gathered from spatial distributions in the environment. Hence, the results of such non-spatial models cannot explain the spatial patterns exhibited in the systems~\cite{muller2022echo, djurdjevac2022feedback}.
\par
\noindent
Contrarily, \newerChanges{other models consider scenarios in which agents are embedded in space and influenced by an external environmental signal, either in a  fixed `cell', or grid-like environment~\cite{baumgaertner2016opinion, baumgaertner2018spatial}, or a jump-like influence from external information~\cite{majmudar2020voter}. } We assume that the information agents receive is defined by an information landscape, for example, emitted by one or multiple information sources, and the agent's position determines the information available to them, hence shaping their opinion. 
We already know that the position of agents in the network matters~\cite{mengers2024leveraging}, but so does the position within the information landscape. 
Compared to other discrete models of initial opinions in homophilic networks, the information landscape can represent continuous information with arbitrary distributions.
\newerChanges{This landscape represents external information available in space, not directly determining internal opinions. Agents sample it locally and integrate it with social input to form their opinion.}
%
%
%
\par
Given the agents in our model are embedded in an information landscape, \newerChanges{we define neighborhood as the local proximity of agents, determined by a finite communication range. While not identical, this definition is conceptually comparable to the thresholds in bounded confidence models of opinion dynamics, where only sufficiently similar agents influence each other~\cite{hegselmann2002opinion, wu_mixed_2023}. However, here the threshold is implicit, emerging from spatial distance instead of explicit opinion similarity, and is decoupled through its embedding in physical space.}
\par
\noindent
To implement homophily in this setting, we require new rules that considers spatiality of the network. In traditional, non-spatial models, homophily can affect links between any pair of nodes, regardless of their spatial distance~\cite{holme2006nonequilibrium}.
However, in spatial systems, interactions are inherently local and limited by proximity. 
Therefore, we propose a potential function based on the dissonance value of agents' opinions, where homophily acts as a local, gravitational pulling force, driving agents to actively move and seek sources of information that minimize this disagreement value. This way of modeling homophily is in line with the aforementioned functional purpose of homophily--reducing disagreement \cite{tornberg2018echo}. 
\newerChanges{In abstract models of opinion dynamics, such as bounded confidence models, it is well known that small confidence bounds can result in echo chambers formation. In this work, we raise a new question in spatially embedded setting: does limiting communication range lead to similar fragmentation, or does the added spatial structure change the nature of consensus formation? Here, we extend that framework to explore this question and test the conditions under which spatial fragmentation occurs, and whether simple decentralized mechanisms can mitigate it.
}
\par
\noindent
\newChanges{Taken together our approach allows us to study emergent phenomena originating from the interplay between spatial dynamics in an (information) landscape, and the distributed opinion formation process. While we model a specific collective estimation scenario, our explicit consideration of agent movements and interactions in space, with their simultaneous impact on the information input
, and the evolution of the network, highlights the functional consequences of spatio-temporal self-organization that is difficult to account for in abstract adaptive network models.} 
\begin{figure}[!t]
    \centering
    \includegraphics[width=1\linewidth]{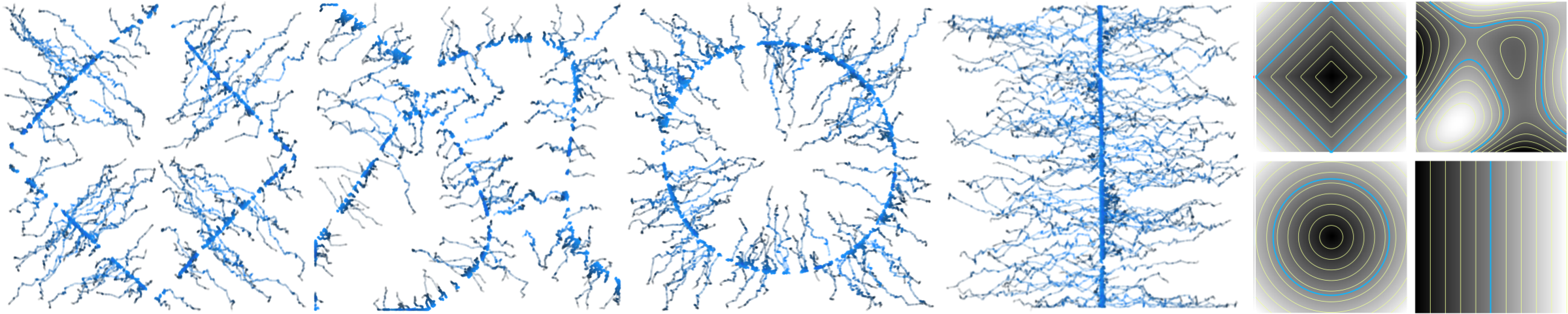}
    \caption{Collective contour capturing emerges as the result of collective opinion dynamics and homophily in space. The traces of agents start with black and become blue (and brighter) over time. Agents converge from the initial random distribution to the mean contour lines of the environment for four different information landscapes. Each plot on the left side corresponds to an information distribution on the right grayscale plot, with the iso-contour lines shown in light yellow and the mean contour line (ground truth) in thick blue line. The video of agents optimizing dissonance function following homophily is available via this link: \url{\homophilyVideoUrl}~\cite{Raoufi2025_vid_dissonance}}
    \label{fig:contour_capturing}
\end{figure}

\par
\newerChanges{
Once clusters form, the reinforcing dynamics of echo chambers may confine agents to their local clusters, limiting their access to diverse information. Here, we ask whether introducing agents with different interaction rules can help reconnect disconnected subgroups and restore collective consensus.
}
\newerChanges{Previous research has shown that stubborn (a.k.a. zealot) agents can regulate the formation of echo chambers~\cite{ghaderi2014opinion, abrahamsson2019opinion, botte2022clustering, muller2022echo}.}
Inspired by this idea, we propose a new role for agents in our spatial setting: Messengers. Compared to the abstract models of stubborn agents, Messengers are active in space.
These agents serve effectively as data ferries that connect sub-populations across wider distances beyond the communication range limits.
Thus, Messengers stand in contrast to the ordinary `Exploiter' state, in which agents continuously update their opinions by optimizing for both conformity and homophily. Messenger agents, by contrast, move freely in space and share their current opinions with others, but do not acquire any new information, i.e., they neither integrate the opinion of their neighbors nor sample the information landscape. 
\par
\noindent
We investigate if this approach can increase the effective communication range and restore collective consensus. A challenge, however, is to define a decentralized mechanism that determines whether, when, and for how long agents remain in the Messenger state. Here, we propose a simple decentralized solution based on the Dichotomous Markov Process (DMP) \cite{bena2006dichotomous}. We study the effect of DMP parameters on individual and collective behaviors, measured by two properties: the ratio of Messengers, and the switching speed between the two states. \newerChanges{By studying the phase diagram of the collective behavior, we aim to identify the characteristic properties of regimes that support consensus. In particular, we investigate how minimal heterogeneity of roles modulates the system's ability to break echo chambers.} %
%
%
%

\section{\newerChanges{Methods}}
In the following subsections, we provide details, first on our agent-based modeling approach to simulate opinion dynamics of collectives with homophilic interactions in space. Then, we introduce the metrics we used to evaluate the behavior and performance of the collective. Finally, we provide information on the parameters used in our simulation. 
\newerChanges{
\subsection*{Modeling Assumptions and Design Choices}
\label{sec:mod_choi}

We consider a finite set of $ N $ agents indexed by $ i \in \{1, \dots, N\} $, evolving over discrete time steps $ t \in \mathbb{N} $. Each agent occupies a spatial position $\mathbf{x}_i^t \in \Omega \subset \mathbb{R}^2 $, where $\Omega$ is a compact, bounded (two-dimensional) arena with reflective boundary conditions that prevent agents from leaving the domain. Agents move within that bounded rectangular arena $\Omega \subset \mathbb{R}^2$, and we apply reflective boundary conditions to ensure that agents remain within the domain. Time evolves in discrete steps, and at each time step, agents update both their opinions and spatial locations. The opinion of agent~\( i \) at time \( t \) is denoted by $z_i^t \in \mathbb{R}$. Communication is spatially constrained by the communication range $r_{\text{comm}}$: agent \( i \)'s interaction neighborhood~$\mathbb{N}_i^t$ includes all other agents \( j \) such that $ \|\mathbf{x}_i^t - \mathbf{x}_j^t\| \leq r_{\text{comm}}$. The environment contains a fixed scalar field f: $\Omega \rightarrow \mathbb{R}$, which defines the spatially distributed information landscape. Agents access this field through noisy local sampling: the environmental signal available to agent $i$ at time $t$ is $s_i^t = f(\mathbf{x}_i^t) + \xi_i^t$, where $\xi_i^t \sim \mathcal{N}(0, \sigma^2)$. 
While the function~\( f \) maps into \( \mathbb{R} \), it is bounded over the compact domain \( \Omega \). This setup reflects our focus on estimation tasks, where opinions represent real-valued quantities drawn from a spatially distributed source. The model remains compatible with bounded opinion spaces when required.
All agents share the same rules of motion and opinion update unless stated otherwise.

Building on this formal setup and our previous work~\cite{raoufi2021speed,ICRA2023Raoufi}, we designed the following modeling components to reflect the spatial and informational constraints of natural and artificial collective systems:
\begin{itemize}
    \item \textbf{Spatially limited interaction network:}
    Our model constrains interactions by physical space and a limited range, resembling the spatial limitations observed in both natural and artificial collectives. In biological systems, such as animal groups, proximity significantly affects interaction likelihood due to sensory or cognitive constraints. In artificial systems, such as robotics or sensor networks, communication is often restricted by signal range. 

    \item \textbf{Spatially continuous environment:} 
    The spatial distribution of information in our model is represented in a uni-modal, radial environment. We chose a continuous environment because it reflects many natural and artificial systems. This choice simplifies the analysis while maintaining ecological and practical relevance. While we demonstrate the generalization of our results to other environments, this foundational framework enables systematic exploration of more complex, rugged, or multi-modal landscapes in future studies. In such scenarios, we anticipate additional behaviors emerging, such as localized echo chambers shaped by environmental topology.

    \item \textbf{Pseudo-gradient descent movement:}
    Agents in our model follow a pseudo-gradient descent strategy to move in space in order to minimize the dissonance value as a potential field. Our previous work proved that even a minimal implementation of gradient-following behavior, such as a phototaxis achievable with simple mobile robots (e.g.,  Kilobot), is sufficient to produce the observed emergent patterns~\cite{rubenstein2012kilobot}.
    
    \item \textbf{Random walk of Messengers:}
    Messengers are agents that stop updating their opinions and act as mobile information carriers (described in more detail in the following). They traverse the environment via a stochastic random walk, simulating exploratory information carriers. We chose this for its simplicity and ability to capture exploratory behavior. Alternative movement patterns, such as ballistic motion or deliberate navigation strategies, would ultimately serve the same purpose and facilitate the diffusion of information over greater distances. This reinforces the robustness of our approach, as the specific choice of movement mechanism does not fundamentally alter the emergent diffusion dynamics.

\end{itemize}
}
In the rest of the section, we first present the components of the baseline model that simulates Exploiter agents, whose behavior is determined by two factors: conformity and homophily (see also \cite{ICRA2023Raoufi}). Then, we introduce a new behavior role for agents, which we refer to as Messenger, with the switching of agents between the two states implemented via a DMP.
\\
\subsection*{Conformity as an opinion-updating rule}
Agents are interconnected via the communication network and continuously exchange information with their local neighbors. Conformity, as a form of social influence, causes individuals to adapt their opinions to reduce their disagreement with others, when they are exposed to their social neighbors' opinions. Therefore, conformity poses a constraint on the opinion of agents in the network. Following the DeGroot social learning model~\cite{degroot1974reaching}, this dynamic constraint describes how agents modify their opinions based on the information they receive from their neighbors. The updating rule of the opinion is formulated as a weighted average of three different components~\cite{raoufi2021speed}: private memory of opinion $(z_{i}^{t} \in \mathbb{R})$, environmental signal $(s_{i}^{t})$, and social signal $(\sum\limits_{j \in \mathbb{N}_i}{{z}_{j}^{t}})$ which are described in: 
\begin{align}
& z_{i}^{t+1} = \alpha  z_{i}^{t} + \frac{1-\alpha}{1 + N_i} \left (  s_{i}^{t} +  \sum\limits_{j \in \mathbb{N}_i} {{z}_{j}^{t}} \right ) \ .
\label{eq:op_update}
\end{align}
The weighted average of these three components shapes the opinion of each individual. The weights are defined explicitly by the self-weight ($\alpha$), and implicitly by the size of the $i$-th agent's neighbor set $N_i = |\mathbb{N}_i|$. The scalar environmental signal $s_{i}^{t}$ is derived from a function $f$ that represents the information landscape. 

\subsection*{Homophily as a motion constraint}
In our model, agents are not fixed in the information landscape but actively move in it to search for sources of information matching their opinions. Similarly, homophily can be seen as the effort to move and find neighbors that match the information the agent receives. This movement is determined by the information agents receive from the environment and their local neighbors, therefore their neighbors can indirectly induce their movement in space. This adds to the formation of opinions and drives more complex collective motions in space. To model this, we used homophily as a mechanism for agents to change their position in space so that the information they receive fits better with the average of their local neighbors. This movement is considered an extra step to increase consensus and is performed in the spatial domain. To implement this movement, we defined an objective function based on the difference of two signals: what the input from the environment is, and what the local neighbors agree upon. The difference generates a potential-like function in space that biases the agents to move to specific points where the difference is minimal. So, homophily is an effort to minimize the \textit{dissonance} as the difference between two values: 
\begin{equation}
    d_{i}^{t} = \frac{1}{2} \left ( s_{i}^{t} - z_{\text{loc},i}^{t} \right ) ^2\
    \label{eq:ObjFunc}\ ,
\end{equation}
with $ z_{ \text{loc}, i} = \sum_{i=1}^{\mathbb{N}_i} z_i / N_i$ being the local collective average that agent $i$ observes.
Agents need to optimize this objective function to satisfy the homophily constraint. To implement it in a distributed way, we applied a minimal sample-wise pseudo-gradient descent (same as in~\cite{raoufi2021speed}), where agents use the differentiation of the samples they measure, as an approximation of the gradient. We used this optimization method since it is independent of the gradient of the objective function, and requires minimal capabilities, being applicable for engineering cases as we showed in~\cite{ICRA2023Raoufi}.
Agents constantly evaluate their dissonance value while they move in space. To approximate the slope of the function at position $\textbf{s}_i^{t}= \begin{bmatrix} x_i^t , y_i^t \end{bmatrix}^\text{T}$, agents calculate the difference of the objective function over the step they took in the last time step. A decaying memory (weighted by $\beta$) of this differentiation smoothens the approximation of the gradient: 
\begin{align}
    & {\nabla}_\textbf{s}d_i^t = \beta {\nabla}_\textbf{s} d_i^{t-1} + (1-\beta) \begin{bmatrix} \frac{\partial d_i^t}{\partial x_i^t},  \frac{\partial d_i^t}{\partial y_i^t} \end{bmatrix}^\text{T}\ ,  \\
    & \frac{\partial d_i^t}{\partial x_i^t} \approx  \frac{\Delta d_i^t } {{x}_i^{t} - {x}_i^{t-1}}, \quad
    \frac{\partial d_i^t}{\partial y_i^t} \approx  \frac{\Delta d_i^t } {{y}_i^{t} - {y}_i^{t-1}}\ .
\end{align} 
To add randomness to the movement of agents, we define a vector along this gradient in addition to a random walk component:
\begin{equation}
    \pmb{\lambda}_i^{t} = (1-\text{r}_\lambda)\frac{-\nabla_\textbf{s}d_i^t}{ \| \nabla_\textbf{s}d_i^t \| } + \text{r}_\lambda \pmb{\eta}_i^t  ,
\end{equation}
in which, $\text{r}_\lambda$ and $ \pmb{\eta}_i^t$ are the weight of the random walk, and a vector of uniform random variables in $[-1,+1]$, respectively. Based on this vector, agents take a step ($\mathbf{w}$) with a fixed size $\text{w}$:
\begin{equation}
     \mathbf{w} = \text{w} \frac{\pmb{\lambda}_i^{t}}{\| \pmb{\lambda}_i^{t} \|}, 
     \label{eq:stepSize}
\end{equation}
In cases where an agent does not have any neighbors, the movement follows only the random walk. Solitary agents will continue walking randomly and update their opinions based on the environmental information until they encounter a neighbor. 
\subsection*{Model extension: Data Ferrying by Messengers}
\label{sec:NewBeh_Msngr}
A potential solution to tackle over-exploitation and the formation of echo chambers due to the limited communication range is to restore the effective connectivity of the network, especially the inter-cluster links of the network. The information can flow across clusters with different opinions and diffuse throughout the network. From an engineering perspective, a trivial solution would be to scale the problem by increasing the communication range, hence pushing the system into regimes with higher network connectivity. Improving the hardware, if possible, comes with physical constraints and increases the cost of the designed system. However, by harnessing the mobility of the agents, an alternative solution is to transmit information via the physical movement of agents carrying it in space. This way, different clusters can exchange information over distances that are possibly much larger than the communication range.
To achieve this, we introduce a new, so-called `Messenger' state for agents to transfer the information as they move in space. A~Messenger can be seen as embodied data that moves around in space and shares the information with its local neighbors that it encounters on its way. This is a similar concept as Zealots or stubborn agents who do not change their opinions~\cite{galam2007role,mobilia2007role,verma2014impact}. A~Messenger moves in space and establishes new links with other agents. The Messenger state should have the following fundamental properties:
\begin{itemize}
    \item A Messenger does not modify its opinion while carrying it around. The value of the data is set to the last opinion of the agent, before becoming a Messenger.
    \item A Messenger moves randomly and independently of environment measurements, the data it carries, or social signals received from its local neighbors.
    \item A Messenger continually shares its fixed opinion value with the others it encounters.
\end{itemize}
In other words, a Messenger agent establishes long-distance uni-directional links by broadcasting its opinion to local neighbors while moving randomly in space. A~Messenger migrating by chance from one cluster to another resembles and implements long, weak ties in a dynamic spatial network~\cite{granovetter1973strength}. The~Messenger state contrasts the ``Exploiter" state. An Exploiter \textit{integrates} the information following the model we explained in the previous subsections, whereas a Messenger behaves as a moving memory of \textit{information}. A receiving agent, whether Messenger or Exploiter, does not distinguish the transmitter of the incoming information. An Exploiter receives social information from other Exploiters or Messengers and integrates it regardless of its source. In contrast, Messengers do not process the incoming information and can be regarded as mobile carriers of information; the information about the history of opinions. In this regard, a Messenger provides a link to the history of its opinion or, in some cases, the opinion of its cluster. 
\par 
Agents can switch back and forth between the two states. When an Exploiter turns into a Messenger, it carries its latest opinion. If this opinion corresponds to the local consensus, then the data represents the opinion of the cluster. Similarly, a Messenger can switch back to an Exploiter state. In that case, the agent forgets its data and replaces it with the current information it receives from the environment. This causes another functional benefit of Messengers which is increased exploration. 
\subsection*{Switching between states: Dichotomous Markov Process (DMP)}
\label{subsect:DMP}
\begin{figure}[!b]
    \centering
    \includegraphics[width=0.5\linewidth]{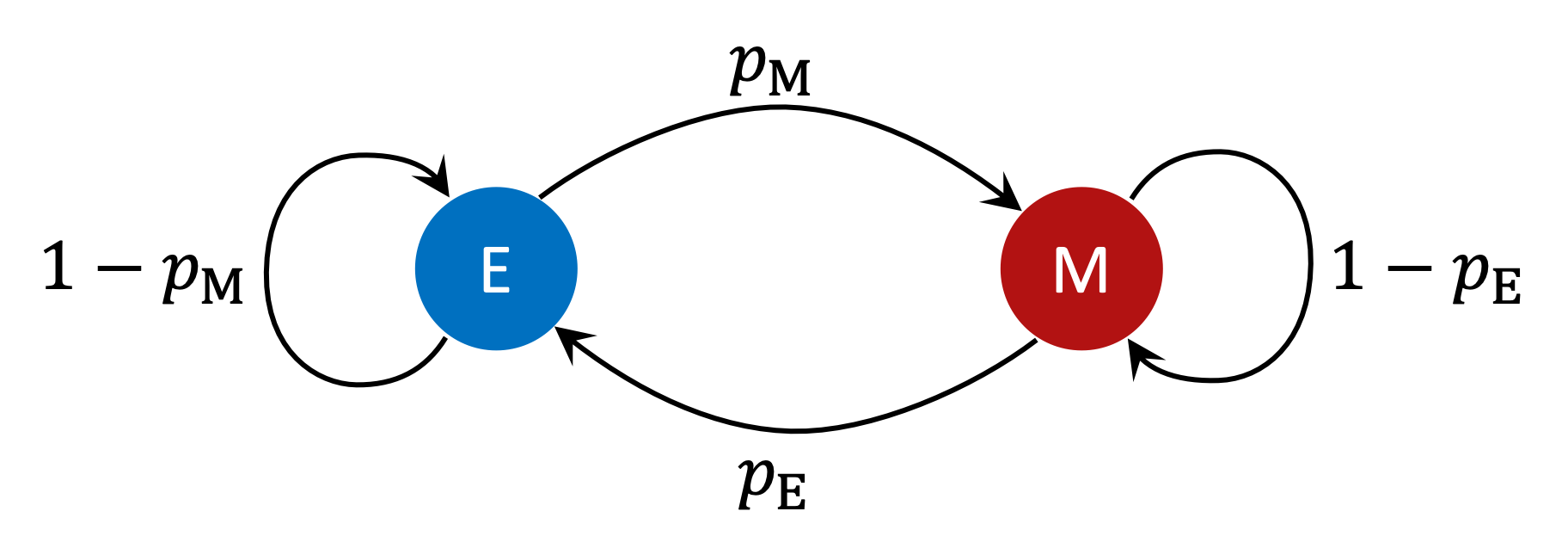}
    \caption{DMP state-machine diagram illustrating how each individual switches its state between Exploiter and Messenger.} 
    \label{fig:dmp_state_machine}
\end{figure}
This specific type of Markov process provides a probabilistic mechanism for switching between two contrasting states, Exploiter and Messenger in our case. It can be easily implemented in a decentralized way and requires only minimal computational capabilities per agent. We implement the DMP such that each agent $i$ at time step $t$ either remains in its current state or switches to the other state. An agent in an Exploiter state switches to Messenger with a transition probability~$p_\text{M}$, and vice versa with probability~$p_\text{E}$, per time step. The complement of each transition probability ($1-p_\text{M}$, and $1-p_\text{E}$) determines the probability of remaining in the current state. The process is illustrated as a state machine in Fig.~\ref{fig:dmp_state_machine}. Note, in general, the DMP is formulated in continuous time in terms of transition rates. The probability of switching per time step is then simply the product of the corresponding transition rate and the time step. We implement the process in a discrete-time domain with $\Delta t = 1$. Hence, we formulate the DMP directly in terms of transition probabilities instead of transition rates~\cite{bena2006dichotomous}. The properties of the behavior for each agent are defined by the two parameters ~$p_\text{M}$, ~$p_\text{E}$. Two properties that are of interest in this paper are the sojourn time and the expected ratio of Messengers. Assuming the stationary process, the sojourn time~\cite{rubino1989sojourn}, denoted as the time between two consecutive switches, for each of the states is defined as follows:
\begin{equation*}
    \mathbb{E}[\tau_\text{M}] = \frac{1}{p_\text{E}}, \quad
    \mathbb{E}[\tau_\text{E}] = \frac{1}{p_\text{M}}.
    \label{eq:DMP_sojourn_times_E_n_M}
\end{equation*}
Then, we define average sojourn time ($\tau_\text{S}$), as the average time spent in either of the states before switching, as the average of the two sojourn times:
\begin{equation}
    \mathbb{E}[\tau_\text{S}]
    = \frac{1}{2} \frac{p_\text{E} + p_\text{M}}{p_\text{E} p_\text{M}} \ .
    \label{eq:DMP_sojourn_time_sw}
\end{equation}
\par
In this paper, we assume all agents have identical parameters. Nonetheless, due to the stochastic nature of the process, the collective properties are probabilistic. Therefore, we provide the relation between the \textit{expected} values of the collective properties and the parameters of the DMP.
The ratio of times agents spend in either of the states determines the ratio of Messengers (and Exploiters) in the collective at a given time step. 
The expected steady-state Messenger ratio ($m$) follows the equation below:
\begin{equation}
    \mathbb{E}[m] = \mathbb{E}\left[\frac{M}{\text{N}}\right] = \frac{p_\text{M}}{p_\text{E} + p_\text{M}}.
    \label{eq:DMP_Mess_ratio}
\end{equation}
\newcommand\figAHeight{0.22}
\newcommand\figBHeight{0.35}
\begin{figure}[!b]
\centering
\includegraphics[width=0.8\linewidth]{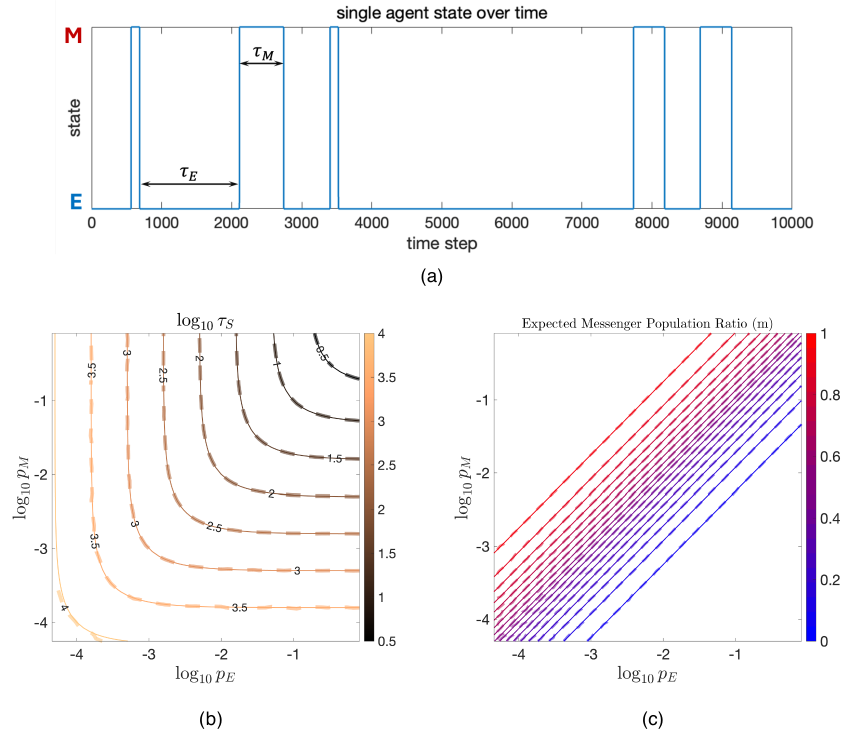}
\caption{Changing the pair parameters of the DMP affects the individual and collective properties. \textbf{a)} An example realization of the DMP with~$p_\text{E~}=~0.003$,~$p_\text{M}~=~0.0004$, showing the time history of a single agent state switching between Exploiter (E) and Messenger (M) states determined by the Markov process. The time duration of staying in either of the states in the $2^{nd}$ sojourn is denoted by $\tau^2_\text{E}$ and  $\tau^2_\text{M}$. \textbf{b)} Average sojourn time of the DMP, and \textbf{c)}~the ratio of Messenger population shown with 5 percent intervals for the analytical and numerical simulations with solid and dashed lines respectively.}
\label{fig:dmp_props_single_exp}
\end{figure}
\subsection*{Metrics and Setup}
\label{sec:Metrics}
Here we define evaluation metrics to benchmark the collective performance. We have metrics of two types: the first type evaluates the collective in the opinion domain, i.e., by using the internal opinion of each agent; while the second does not require access to the internal state of the agents, but their position in the physical space. By quantifying the precision of the collective opinion, the spatial arrangement of agents in space, and the connectivity of the communication network, these metrics capture all relevant aspects of the collective performance. To show how Messengers change the performance of the collective, we normalize the absolute metrics to that of the \textit{baseline} setup, where there is no Messenger in the collective. Please note that in our finite-size stochastic simulations, the expected ratio of Messengers for the bottom right corner of the parameter space is $\mathbb{E}[m] < 10^{-8}$, which validates the zero-Messenger assumption. We refer to these metrics as normalized metrics. 

\par
Precision error is the first metric we used. Since the objective of the problem is similar to~\cite{raoufi2021speed}, we use the same metrics and refer readers to the original paper for further details. We decomposed the total accuracy error into trueness and precision errors. In this paper, we assume that the initial distribution of agents is diverse enough and the trueness error (collective bias) is negligible (${z}_\text{col} \approx z_\text{gt}$). Hence, we only report the precision error as it contains the necessary information about achieving consensus. We obtain the \textit{opinion} precision error as the variance of the opinions with respect to the collective (average) opinion (${z}_\text{col}=\sum_{i=1}^\text{N} {z}_i / \text{N}$) as following:
\begin{equation}
    E_\text{P}^\text{O} = \frac{1}{\text{N}} \sum\limits_{i=1}^\text{N}({z}_i - {z}_\text{col})^2 . 
\end{equation}
To show the precision of the spatial positioning of the collective, i.e., spatial consensus, we use the same metric and take the position of agents instead. This metric directly quantifies the performance in terms of contour capturing~\cite{ICRA2023Raoufi}. 
The precision error in the \textit{space} domain measures how close are agents to the collective average with regard to the information distribution. We distinguish the two domains by a superscript and define the precision error in the \textit{space} domain as:
\begin{equation}
    E_\text{P}^\text{S} = \frac{1}{\text{N}} \sum\limits_{i=1}^\text{N}({{s}}_i - {{s}}_\text{col})^2 . 
    \label{Eq:E_P_S}
\end{equation}
\par
Number of clusters is another metric that measures the cohesion of the agents in the physical space, particularly considering their communication network. 
To quantify the number of clusters, we identify and count the connected components of the spatial communication network, determined by the communication range of agents.
\subsection*{Simulation Configuration}
As an extension of our previous work~\cite{raoufi2021speed}, we used the same parameters, except for the communication range.
To put the system in a regime where echo chambers can potentially emerge, we set the communication range to $r_\text{comm}=0.15$. This is similar to the configuration of Kilobots~\cite{rubenstein2012kilobot}, a robotic platform used to study collective behaviors, which we used to implement the baseline setup in a real-world setting~\cite{ICRA2023Raoufi}. \newerChanges{The environmental signal $ s_i^t $ is treated with equal weight as each neighbor ($z_j^t$) to ensure that external and social influences evolve on comparable time scales.} 
Also, we distributed the probabilities ($p_\text{E}$, $p_\text{M}$) exponentially, same as in Fig.~\ref{fig:dmp_props_single_exp}. This way, we can reveal a wide spectrum of the DMP dynamics on various scales. The results that we report are the average of 24 independent Monte-Carlo simulations. Otherwise noted, we used the parameters of the model as reported in Table~\ref{Table:params}.
\begin{table}[!ht]
\begin{tabular}{ |c|c|c|c| } 
\hline
Name                & Description & Value \\
\hline
$\text{N}$          & Number of Agents                                  & 100 \\ 
$A$                 & Arena Size                                        & $2\times2$ \\
$r_\text{comm}$     & Communication Range                               & $0.15$ \\ 
$w$                 & Walking Step Size                                 & $0.002$ \\ 
$t_\text{f}$        & Simulation Time Step Duration                     & 50,000 \\
$\sigma$            & Measurement Noise                                 & 0.001 \\
$\delta_{t}$        & Integration Interval                              & 1 \\ 
$\alpha$	        & Self-weight on Opinion Memory                     & 0.99 		 \\ 
$\beta$	            & Decaying Factor for Gradient Descent 	            & 0.5 		 \\ 
$\text{r}_\lambda$  & Random Walk scalar                                & 0.001 \\
$p_\text{E}$        & Probability of Switching to Messenger State       & $\exp{\{-20:-2\}}$ \\
$p_\text{M}$        & Probability of Switching to Exploiter State       & $\exp{\{-20:-2\}}$ \\
\hline
\end{tabular}
\caption{Simulation Parameters}
\label{Table:params}
\end{table}
%
\section{Results}
\label{sec:Results}
In this section, first, we review consensus formation in the baseline model with only Exploiter agents and study the emergence of echo chambers in low connectivity regimes. We then investigate the ability to restore consensus via the introduction of an additional state of individuals, which we refer to as Messenger, and a DMP governing the stochastic switching between the Exploiter and the Messenger states. The model details are given in the Methods section~\ref{subsec:Model}.
\subsection*{Dissensus in Space: The Emergence of Echo Chambers} 
\begin{figure}[!t]
    \centering
    \includegraphics[width=0.7\linewidth]{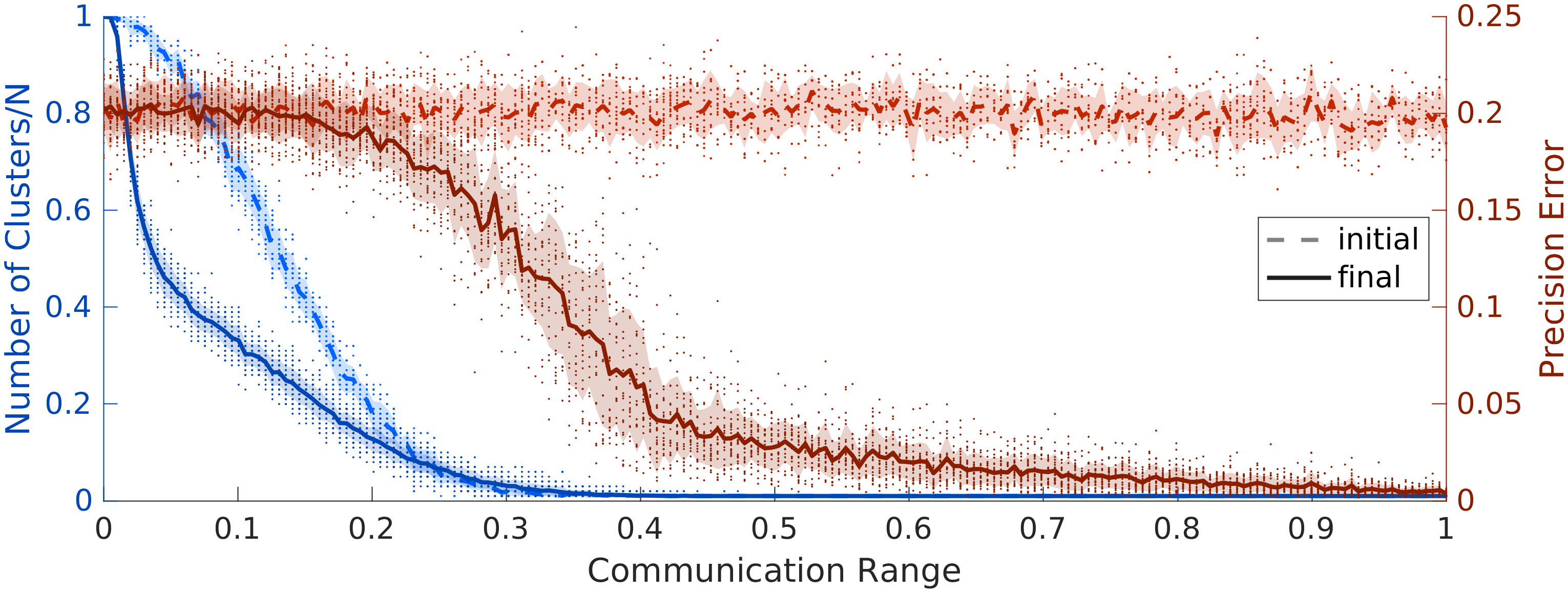}
    \caption{Normalized number of clusters and opinion precision error decrease by communication range at the beginning and end of simulations for the \textit{baseline} setup without Messengers. The information landscape shown here is a radial distribution centered in the arena. 
    }
    \label{fig:nCluster_vs_comRange}
\end{figure}
As shown previously in~\cite{raoufi2021speed}, the coupling of the local information aggregation and homophily not only allows agents to estimate the mean value of information distribution in the environment but also leads to an emergent spatial collective pattern formation in the information landscape (see Fig.~\ref{fig:contour_capturing}.) We refer to this spatial consensus as collective contour-capturing behavior, i.e., the aggregation of agents at the average iso-lines of the information landscape. The resulting patterns show that in seeking specific information sources, individuals actively move in space, and form communities based on the information they receive. 
Here, we explore the result of such complex behavior in more detail and show that global convergence relies on a sufficiently large communication range. In settings with reduced network connectivity, however, global consensus breaks down, and only local consensus can be observed.
Due to reduced network connectivity and homophily, the system becomes trapped into local minima, leading to the formation of clusters that serve as echo chambers, with negative consequences on the collective performance due to the inhibition of information flow across the clusters with dissimilar opinions (see Fig.~\ref{fig:nCluster_vs_comRange}). 
\par
\begin{figure}[!b]
\centering
    \includegraphics[width=0.7\linewidth]{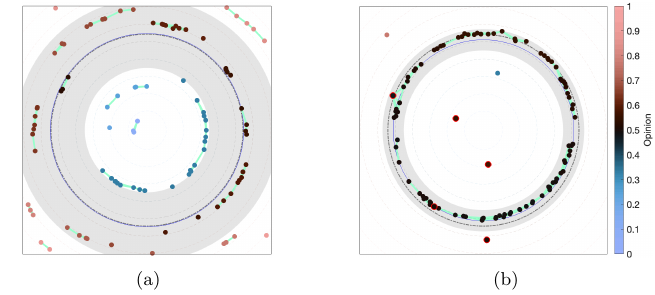}
    
\caption{Final position of agents in the information landscape. \textbf{a)} Formation of echo chambers due to limited connectivity and homophily (baseline setup), \textbf{b)} consensus is achieved by introducing Messengers (with red circles around them). The fill-in colors correspond to the opinion agents have, the green lines show the links of the communication network, and the width of the gray ring shade indicates the spatial precision error. The dashed gray rings and the blue ring show the contours of the information distribution, and the contour related to the mean value of the collective, respectively. 
}
\label{fig:echo_chambers_emrg}
\end{figure}

Our results demonstrate that these clusters inhibit collective movement and prevent the system from achieving consensus in space, i.e., the collective contour-capturing. The clusters form partial contours each aligned locally with the iso-line of the information landscape corresponding to the within-cluster estimation of the average value of the information landscape. 
An example of such cluster formation is shown in Fig.~\ref{fig:echo_chambers_emrg}-a, where the echo chambers appear as concentric arcs of various radii, in this case for radial information distribution; while in Fig.~\ref{fig:echo_chambers_emrg}-b we see a consensus pattern, where agents form a single contour at the average value of the information landscape.   
\par
To quantify how the consensus, and the resulting contour-capturing behavior, rely on sufficient connectivity, we measured the number of clusters being formed as well as the precision of consensus ($E_\text{P}^\text{S}$, for further information, see Sec.~\ref{sec:Metrics}) at the end of the experiment.
A consensus \textit{in space} is achieved when agents are precisely on the same contour line. We illustrate these two metrics in Fig.~\ref{fig:nCluster_vs_comRange}. The number of clusters and precision error decrease with the communication range after a critical value ($r_\text{comm}\approx0.3$), highlighting a specific regime where the consensus breaks due to the insufficient network connectivity. This critical value is the minimum required for the initial network to be fully connected. 
By comparing the final and initial state of the collective in Fig.~\ref{fig:nCluster_vs_comRange}, we observe that the basic method still improves both the cohesion and precision of the collective. In the next sections, we propose and investigate a solution for increasing effective connectivity by utilizing information-carrying mobile agents, `Messengers',
with the switching in and out of the Messenger state for individual agents being governed by the DMP.

\subsection*{Dichotomous Markov Process as a State Switching Mechanism}
\newcommand\figTwoHeight{1.7}
\newcommand\figTwoWidth{1.5}
\newcommand\figTwoWidthDouble{5}
DMP is a simple stochastic process with the switching dynamics defined by only two parameters $p_\text{M}$ and $p_\text{E}$. Fig.~\ref{fig:dmp_props_single_exp}-a illustrates the temporal dynamics of a single agent's state. Changing the parameter pair influences two properties of a single agent behavior: the ratio of time each agent spends in either of the two states; and how quickly they switch between the two states. The latter, known as the sojourn time (see Sec.~\ref{subsect:DMP}), refers to the time between two consecutive switches. We illustrate the time spent in the Exploiter and Messenger states in Fig.~\ref{fig:dmp_props_single_exp}-a by $\tau_\text{M}$ and $\tau_\text{E}$, respectively. At the collective level, changing the two parameters affects the collective average sojourn time ($\tau_\text{S}$), and the ratio of Messengers~($m$). These two properties are depicted in the 2D parameter space in Fig.~\ref{fig:dmp_props_single_exp}-b, and Fig.~\ref{fig:dmp_props_single_exp}-c.
\par
We evaluated the Markov process by running multiple independent Monte-Carlo simulations and validated it by comparing the numerical simulation results to the known analytical solution given in Eq.~\ref{eq:DMP_sojourn_time_sw} (also shown in Fig.~\ref{fig:dmp_props_single_exp}-b, and Fig.~\ref{fig:dmp_props_single_exp}-c). The switching speed reaches its maximum in the top-right corner of the parameter space, where $p_\text{M}$ and $p_\text{E}$ are both large ($\log_{10}p_\text{E} \approx 0$ and $\log_{10}p_\text{M} \approx 0$). It decreases by moving toward the bottom-left corner (small values of $p_\text{E}$ and $p_\text{M}$). In the top-left corner of Fig.~\ref{fig:dmp_props_single_exp}-c, the collective is mainly comprised of the Messengers, and the ratio decreases diagonally toward the bottom-right corner. In the following sections, we will explore how the two DMP parameters determine the collective performance in terms of the consensus in the opinion domain, and consensus in information landscape, i.e., contour-capturing behavior. For the rest of the paper, we will refer to the setup without Messengers as the \textit{baseline} setup.  
\subsection*{Opinion Consensus with Messengers}
The ability to arrive at the consensus for the heterogeneous collective consisting of both Messengers and Exploiters depends on two parameters of the DMP ($p_\text{M}$ and $p_\text{E}$) governing the switching process. A balanced set of properties ($m$ and $\tau_\text{S}$) is required to enable Messengers to propagate information efficiently beyond the limitation of the communication range. This also applies to the Exploiters, which are needed to process the available information in the collective. 
\par
\begin{figure}[!b]
    \centering
    \includegraphics[width=0.55\linewidth]{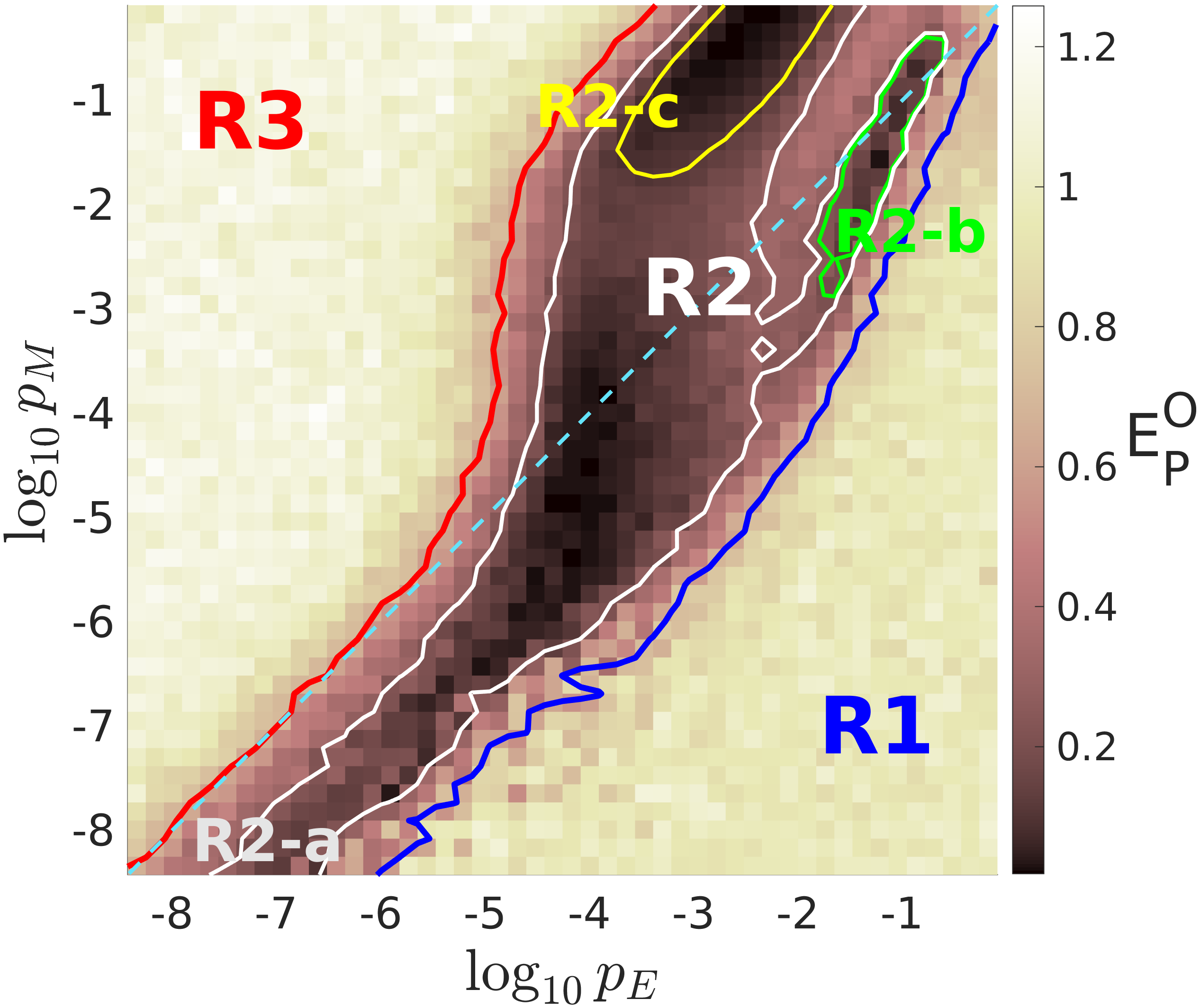}
    \caption{Normalized opinion precision error w.r.t the baseline setup indicating different areas in the parameter space. R1: Too few Messengers (greedy collective behavior); R2: Set of regions with a balanced number of Exploiters and Messengers, R2-a (Specialized Exploiters): Specialized slow-switching agents with Exploiters in the majority, R2-b (Enhanced Exploration): Fast-switching agents, R2-c (Generalized Messengers): Majority being fast-switching Messengers; R3: Too many Messengers. The supplementary video of the simulations for a sample of each region is available via this link: \url{\supplementaryurl}~\cite{Raoufi2025_vid_DMP}.}
    \label{fig:res_dif_regions}
\end{figure}

We illustrate the normalized opinion precision error ($E_\text{P}^\text{O}$) across the parameter space in Fig.~\ref{fig:res_dif_regions}. Each point in the 2D space of $p_\text{M}$ and $p_\text{E}$ represents a pair of parameters corresponding to a specific configuration for the DMP. The bottom-right corner point is analogous to the baseline setup, where the ratio of Messengers is zero. This serves as our reference for normalization, with each point's normalized performance being the ratio of its absolute performance to that of the baseline. The figure shows that the baseline setup (bottom-right corner), where all agents are Exploiters, is not optimal. The higher performance of other regions indicates that introducing the Messengers promotes consensus. Changing the parameters can move the system across different regimes, with certain configurations being locally optimal. The analysis of these local optimal regions increases our understanding of the conditions necessary for achieving consensus and provides guidelines for designing such systems. For ease of reference, we labeled different regions of the parameter space exhibiting qualitatively similar behavior. 
\par
According to the properties of the DMP (see Sec.~\ref{subsect:DMP}), the bottom and left parts of the parameter space correspond to the regions with slow switching of states on average (large $\tau_\text{S}$), with agents remaining in at least one of the states for very long times. Given that agents in these regions rarely switch their states (see Fig.~\ref{fig:dmp_props_single_exp}-b), we can consider them as \textit{specialized} individuals. This is in contrast to the top-right corner, where \textit{generalized} agents frequently switch between the two states, corresponding to short sojourn times. The generalized agents experience both states during the course of the experiment. Additionally, if we divide the space off-diagonally into two triangles, with the lower right triangle dominated by Exploiters. 
\subsection*{Integration-vs-Information Tradeoff} 
In Fig.~\ref{fig:res_dif_regions}, as we go up diagonally from the bottom-right corner to the upper-left corner, we encounter collectives with varying proportions of Messengers. For instance, in region R1, Exploiters constitute the largest share of the population on average resulting in strongly clustered collectives, where agents immediately get trapped in local echo chambers. The opinion consensus performance in this region is deficient due to the prevalence of many Exploiters, which make the collective behavior greedy. We consider an Exploiter agent to integrate and process information according to the DeGroot model for social learning (see Sec.~\ref{subsec:Model}.) Therefore, R1 is dominated by the \textit{integration} of the available information. Their exploitative behavior prevents the flow of new information across clusters. In this setting, the system exhibits the most conservative behavior in terms of mobility and tends to over-exploit in terms of information aggregation. For engineering applications, especially where moving is costly or hazardous, selecting parameters closer to this region might be preferred. Conversely, R3 represents an overly dispersed region where extreme exploration of Messengers and inadequate processing of available information negatively impact performance. In this region, too many Messengers, acting as moving \textit{information}, are redundant; while there are insufficient Exploiters to integrate new information into the collective. In R3, collectives demonstrate the highest network plasticity, meaning any two random agents are likely to meet each other, irrespective of their opinions. R1 and R3 are extremes of either exploitation or exploration, respectively. The high-performing regions are situated between these two extremes, closer to the off-diagonal line (the blue dashed line). We labeled this \emph{balanced} region as R2. This region contains all local optima between the blue and red contours confining R1 and R3, respectively. Next, we examine different locally optimal regions within R2, focusing on varying switching timescales.
\par
\subsection*{Generalization-vs-Specialization Tradeoff} 
Moving from the bottom-left to the top-right corner of the parameter space decreases the timescale (increases the frequency) of switching, as shown in Fig.~\ref{fig:dmp_props_single_exp}-b (see also Eq. \ref{eq:DMP_sojourn_time_sw}). The change in the DMP timescale influences the optimality of regions within R2. Near the bottom-left corner, R2-a is located below the off-diagonal line. The corner represents \textit{specialized} agents, that effectively, due to the low switching probabilities, do not switch their state during the simulation and maintain their initial Exploiter or Messenger states. 
The tail of R2-a aligns with the iso-lines of Messenger ratios (see Fig.~\ref{fig:dmp_props_single_exp}-c). This indicates that the system with static roles performs best for a specific ratio of Messengers that is significantly less than half. Accelerating the switching frequency (towards the top-right) increases the likelihood of agents experiencing both states, thus making them more \textit{generalist}. As we go up in this space, the high-performing points gradually shift towards higher Messenger ratios near the center of the parameter space. This pattern suggests that, on average, more Messengers are needed in collectives of generalists compared to the specialized ones. Increasing the switching frequency further results in the branching of R2 into two distinct sub-regions: R2-b and R2-c. 
\par
R2-b, the lower branch of this fast-switching region, is not parallel to the iso-lines of $m$. This implies that the optimality of R2-b is caused by both properties of the DMP. In this region, the points below the off-diagonal line (lower $m$) are associated with longer timescales. Whereas points with higher Messenger ratios (above the line) favor faster switching. Longer timescales for Messengers mean that they have more time to transport the information they carry, effectively lengthening the link they create. In other words, smaller numbers of Messengers should move for longer durations. However, compared to R2-a, Messengers in R2-b switch back to the Exploiter state much faster. Consequently, R2-b is an optimal region, especially in scenarios where long random movements incur significant costs. Observations of agents' behavior during simulations reveal that this configuration indirectly enhances the collective's exploration behavior. 
\par 
Unlike R2-b, R2-c comprises a majority of Messengers and is elongated parallel to the iso-lines of Messenger ratios. This pattern suggests that the optimal parameter configurations in R2-c primary depend on $m$, and are not significantly sensitive to $\tau_\text{s}$. This region is distinguished by conservative information mixing, promoted by the abundance of Messengers moving randomly in space. 
The slower configurations outside of R2-c are less optimal, as Messengers transport information across longer distances than necessary. The superiority of faster dynamics in this region underscores the significance of spatio-temporal coupling in the problem. We later show that, compared to R2-b, convergence is achieved more rapidly in the R2-c region (see Fig.~\ref{fig:res_temporal_development}.)
\par
\subsection*{Temporal Evolution of Parameter Space}
\begin{figure*}[!t]
\centering
    \includegraphics[width=\linewidth]{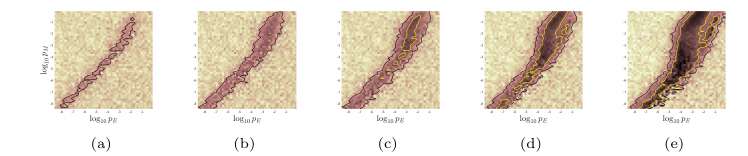}
\caption{Time development of parameter space for precision error. Panels \textbf{(a)}, \textbf{(b)}, \textbf{(c)}, \textbf{(d)}, and \textbf{(e)} correspond to \( t = 1\text{k} \), \( t = 2\text{k} \), \( t = 4\text{k} \), \( t = 10\text{k} \), and \( t = 40\text{k} \), respectively. Bifurcation of the optimal parameters happens for the top-right side of the parameter regime corresponding to fast dynamics, while the bottom-left corner does not undergo a significant shift, showing a more robust performance by the time limit.}
\label{fig:res_temporal_development}
\end{figure*}

By examining snapshots of the parameter space at different time steps, we obtain insights into the temporal evolution of the various regimes in the parameter space. Observing the system's progression over time helps us track the emergence and development of each of the optimal regions (as shown in Fig.~\ref{fig:res_temporal_development}). This approach enables a comparison of the temporal characteristics of each region and allows us to evaluate their performance in scenarios with different time budgets available for the overall task. 
For instance, the high-performing optimal region located at the bottom-left quarter of the space demonstrates consistent performance, irrespective of the time limit. It maintains its optimality across all time limits, suggesting a more robust and reliable performance. This region is characterized by agents that are \textit{specialized} according to their initial states. The gradual slight diagonal shift of this region's tail to the lower right, observable from Fig.~\ref{fig:res_temporal_development}-a to Fig.~\ref{fig:res_temporal_development}-e, indicates that a lower (specialized) Messenger ratio is favorable in scenarios with longer time budgets.
\par
In contrast, the local optimum for generalist Exploiters, located in the top right narrow valley in the precision error (R2-b), emerges only when the system has enough time to process the task. It is important to note that this region is advantageous as it minimizes precision error in both the physical and opinion domains, as we discuss further below in Sec.~\ref{sec:cont_capt}. Additionally, we observe a bifurcation in the fast-switching parameter regime. Specifically, the small head of the elongated R2 (see Fig.~\ref{fig:res_temporal_development}-b) grows over time and eventually branches into two distinct regions (R2-b and R2-c). The gap between the two branches widens over time. 
%
%
Another notable temporal shift occurs in the region with generalist agents and a time-averaged majority of Messengers, the larger upper left branch (R2-c,) which gradually ascends toward higher Messenger ratios. This suggests that in scenarios with higher time budgets, increasing exploration efforts--by assigning more agents to perform random walks--is advantageous. The final aspect to highlight is the continuous improvement in the precision of the optimal regions. The precision gain over time demonstrates the existence of a speed-accuracy tradeoff in this scenario. In contrast, extreme parameter setups, such as those characterized by excessive exploitation (R1) and random diffusion of Messenger (R3), do not exhibit any significant improvement over time. 
\newcommand\figTempDevWidth{0.3}
\subsection*{Contour Capturing Performance}
\label{sec:cont_capt}
We turn now to evaluating the precision error in the spatial domain, measured by~$E_\text{P}^\text{S}$. This metric quantifies the collective performance in contour-capturing tasks and is particularly useful when access to the internal opinion of agents is not possible.  
The results, as depicted in Fig.~\ref{fig:res_init_mes}-a or Fig.~\ref{fig:res_init_mes}-b, do not offer any additional high-performing regions compared to those already identified in Fig.~\ref{fig:res_dif_regions}. However, the local optimal regions above the diagonal disappear. This is because excessive random walks by too many Messengers do not contribute to spatial convergence. The difference between the two metrics, $E_\text{P}^\text{S}$ and $E_\text{P}^\text{O}$, signifies that while consensus may be achieved in the opinion domain, it does not necessarily translate to a spatial consensus.
\par
Unlike opinion consensus, successful contour capturing requires agents to be more conservative. Indeed, the random movement of an excessive number of Messengers in space increases the system's spatial precision error.
Similar to the findings in opinion consensus, the same narrow optimal region representing generalized Exploiters (similar to R2-b in Fig.~\ref{fig:res_dif_regions}), and the long tail characterizing specialized Exploiters (comparable to R2-a) are again identified as high-performing regions in terms of spatial consensus. This observation suggests that converging on similar sources of information promotes the achievement of opinion consensus, but not necessarily vice versa.    
\par 
\subsection*{The Effect of Initial State Distribution of the DMP}
\label{sec:init_DMP}
\begin{figure}[!t]
\centering
\includegraphics[width=0.9 \linewidth]{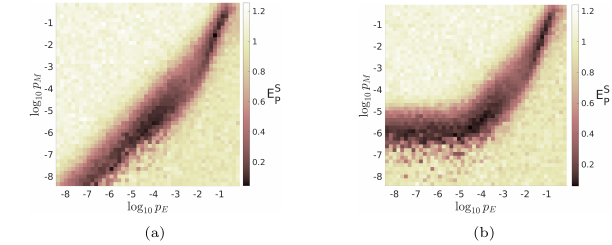}
\caption{Normalized spatial precision error for different parameters of the DMP. The effect of the initial number of Messengers on the contour capturing performance for \textbf{a)} initial state with the expected number of Messengers, \textbf{b)} initial state without any Messengers. }
    \label{fig:res_init_mes}
\end{figure}

We already have established that the temporal properties of the DMP significantly affect collective performance. Another important aspect to consider is the distinction between the DMP transient and stationary behaviors. Up to this point, we have assumed that the DMP begins with the expected analytical ratio of Messengers, thereby initially placing the process in its stationary state. However, any deviation from this stationary state puts the system into a transient phase. 
The progression towards reaching the stationary state is time-dependent and the transient behavior of the DMP also influences the collective performance. Specifically, we show that the transient behavior is linked to the DMP relaxation time ($\tau_\text{C}$) ~\cite{bena2006dichotomous} (see Fig.~S1 and Eq.~S1 in the Supplementary Information). Next, we will explore how the system's behavior is affected by its initial condition, particularly focusing on the initial population of Messengers. 
To show this difference in Fig.~\ref{fig:res_init_mes}, we conduct a comparison to a case where the system starts without any Messengers. We know that, given fixed transition rates, the system requires time to attain the expected analytical stationary properties, namely the Messenger population ratio. 
\newcommand\figInitDistWidth{0.3}
\par
The observed differences are particularly noticeable in DMPs with slow timescales (bottom-left corner in Fig.~\ref{fig:res_init_mes}). A comparison of the two figures shows that for slow-switching dynamics, the tail of the optimal region bends upward under non-stationary initial conditions. This suggests that the absence of initial Messengers is compensated for with higher $p_\text{M}$ values. In such scenarios, the actual time-averaged Messenger ratio is below its expected stationary value due to the transient behavior of the Markov process.

Same as the temporal development studied earlier in Fig.~\ref{fig:res_temporal_development}, by comparing the parameter space at different time steps we observe that systems with different time-scales reach their stationary state with different speeds (see Fig.~S2). For example, systems with fast-switching dynamics in the top-right corner reach their stationary states more quickly, which makes them more robust to varying initial conditions within the same time constraints. In contrast, slower systems require more time to gradually approach their stationary states. Consequently, we observe that the horizontal tail of R2 shifts downward, eventually aligning with a shape similar to that depicted in Fig.~\ref{fig:res_temporal_development}-a.
With these observations, we highlight the importance of accounting for transient effects, particularly in scenarios where the properties of the system must dynamically change from one configuration to another. They also provide insights into the robustness of each configuration against variations of initial conditions and selecting parameters from different optimal regions. 
\subsection*{Different Information Distributions}

\begin{figure}[!t]
    \centering
    \includegraphics[width=0.9\linewidth]{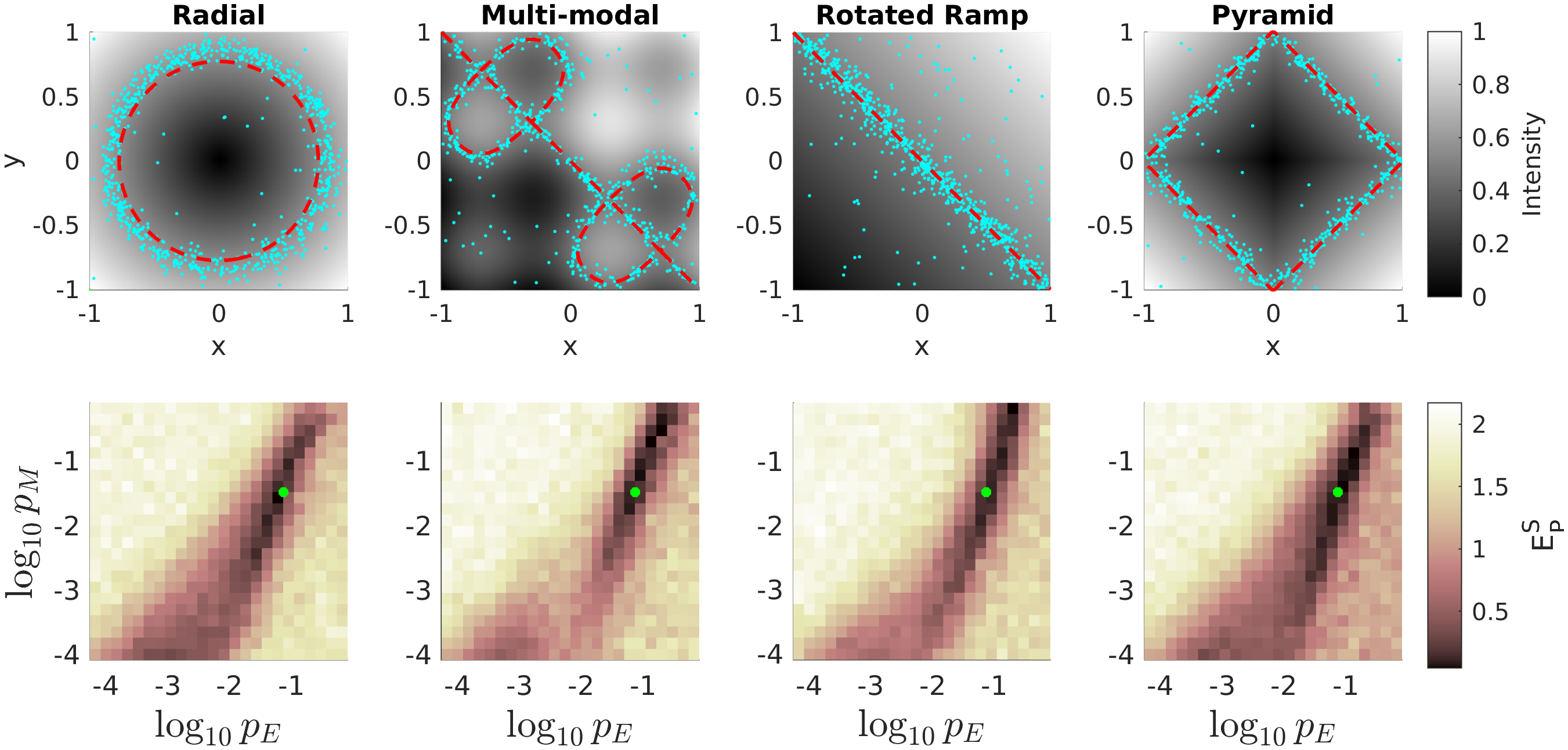}
    \caption{The performance of the collective in different environment distributions. First row) the information landscape, and its mean contour line (red dashed line). The cyan dots show the scattered position of agents at the end of each simulation for the specific parameter marked by the green dot in the respective figure in the second row. Second row) the precision error ($E_\text{P}^\text{S}$) for the corresponding environment.}
    \label{fig:res_dif_envs}
\end{figure}

Our main focus in this paper is on the specific case where the spatial distribution of the information is radial, i.e., the iso-contour lines are circles. This configuration resembles many natural situations where a point-like source isometrically emits a concentration into its surroundings, which may then spread through diffusion or convection.
To ensure generality, we also tested our model's performance in environments with various other distributions. The first row of Fig.~\ref{fig:res_dif_envs} presents different information distributions, where agents (depicted as cyan dots) converge to the zero-bias (ground-truth) contour line (indicated by a red dashed line). The introduction of Messenger with tuned DMP parameters shows an improved performance across all four environmental benchmarks. \newerChanges{We observed that in all four environmental scenarios, the baseline setup (without Messengers) leads to the emergence of echo chambers, indicating that dissensus arises in sparse regimes independent of the specific shape of the information distribution.}
\par 
In the second row of Fig.~\ref{fig:res_dif_envs}, we present the optimal regions within the parameter space for different information distributions. While the shape of these optimal regions varies slightly depending on the distribution, the general features of the error landscape and the underlying mechanisms discussed in previous subsections remain unchanged. Here, we restricted our analysis to the top-right quarter of the parameter space, where the most non-trivial optimal regions are located. 
Furthermore, we selected a specific optimal parameter pair (marked with the green dot) to demonstrate the final agent spatial distribution for this particular DMP configuration. This visualization confirms that the chosen parameter set results in a satisfactory spatial consensus across all tested information distributions. Additionally, we also adapted our model to an abstract 1-dimensional environment, with a few modifications, such as the implementation of the random walk (see Appendix). The corresponding results show only negligible qualitative differences compared to the 2D environment. 
%
%
%
\section{Discussion}
The integration of explicit spatial behavior into the collective opinion dynamics model introduces a new dimension to the behavior's complexity. 
\newChanges{With this new dimension, we emphasize that collective dynamics unfold not only in time but also in \textit{space}. Collective behavior in real-world settings is inherently organized in and through space. Abstracting away the spatial dimension when modeling these systems risks oversimplifying or stripping away relevant aspects of emerging collective behavior. By considering both dimensions, we can capture their interplay and better understand these dynamics. }
\newChanges{Our model demonstrates spatial dynamics through several components: a) initializing opinions based on spatial distribution, b) defining the interaction network according to spatial local proximity, and c) driving agents' movement in space through homophily. The interplay of these elements results in complex collective behavior, where homophily in space leads to emergent patterns that trace features of the spatial distribution of information, resulting in the so-called `contour capturing behavior.'} 
\newChanges{Unlike ad-hoc rules for modeling homophily, such as disproportional rewiring probabilities, our approach shows that seeking reaffirming information sources in space naturally leads to the emergence of echo chambers. 
This eliminates the need to create artificial rules engineered to fit the specific data.}

\newChanges{The insights gained from this modeling approach extend beyond abstract theory with potential applications to engineering or even computational social sciences.} 
\newChanges{On the one hand, understanding the challenges and fail-prone configurations of a system enables the design of a more robust collective robotics system, such as in the application of environmental monitoring or contour capturing behavior as in our previous work~\cite{ICRA2023Raoufi}. 
On the other hand, a clearer understanding of the factors contributing to echo chamber formation can, for example, inform strategies to mitigate polarization. Humans continuously decide which locations, events, or activities physically to attend. Some of these decisions will be driven in part by homophilic interactions, and the attendance will contribute to the individual's information acquisition. However,  real-world human dynamics are undoubtedly more complex that the collective estimation scenario considered here. They occur in more complex and structured spaces and individual opinions are multi-dimensional. Nevertheless, the general conceptual insights from our research remain relevant. 
}

\newChanges{
Recent work in epidemic modeling similarly highlights an increase interest, enabled by technological advances, in moving beyond ad-hoc assumptions on the structure of contact networks, towards using human spatio-temporal mobility data to understand the origin of network structures constraining contagion dynamics~\cite{brockmann2009human,barbosa2018human,chang2021mobility}.  
}
%
The collective pattern formation observed in our model represents a type of spatial consensus, and we show that it is conditional on having sufficient network connectivity. Our simulations for collectives with short communication range indicate that the exploitative mechanisms leads to the collective becoming trapped in local optima, a process we associate with the emergence of echo chambers.
A unique feature of these echo chambers is their similar shapes, which mirror the spatial information distribution. The spatial patterns constructed by the echo chambers reflect the properties of the information landscape and are a consequence of the spatial dynamics of the system, typically ignored in abstract opinion formation models. 
Also, our findings indicate that these echo chambers inhibit the flow of information, which prevents the collective's ability to reach a precise consensus.  It is important to clarify that the impact of communication range is relative to the length scale of agents' distribution. Low connectivity can stem from either a limited communication range or a large initial diversity in opinion space. This observation highlights a potential disadvantage of excessive opinion diversity in a collective when the communication range is limited. 
\par
We suggested that a method to overcome the local traps inherent in low-connectivity networks is to indirectly extend the effective communication range. This is particularly relevant in systems comprising mobile agents. By leveraging mobility and employing certain individuals as embodied data carriers, individuals can interact beyond the physical limits of their communication. To this end, we introduce a new behavioral role for agents, the \textit{Messenger} state. This state allows designated agents to function exclusively as carriers of information. 
A Messenger is a stubborn agent with a fixed opinion, sharing its opinion while moving randomly in space. These agents can be viewed as mobile ``quasi-memory'' of agents' opinions, with their independent random movement promoting broader exploration. 
We employed the Dichotomous Markov Process (DMP) as a simple, distributed switching mechanism between behaviors. The agents switch randomly at each time step whether they should switch to the other state or stay in their current state. A key advantage of this approach is its decentralization, minimal computation requirements, and scalability. 
\par
Still, the two parameters of the DMP add an additional degree of freedom. Modifying the DMP parameters determines the temporal properties of the switching mechanism for a single agent. This, in turn, indirectly modifies the collective's properties, such as the ratio of Messengers and the average speed of switching (denoted as $m$, and $\tau_\text{S}$). The expected value of these two properties can be derived analytically as a function of the DMP parameters. 
We showed that these properties significantly influence the collective performance, in achieving consensus in both the opinion and the spatial domains.
\par
Our numerical results identify several local optimal regions where the collective achieves the highest precision in consensus, which is the ultimate objective in this scenario. Each region exhibits unique characteristics, reflecting a spectrum of collective behaviors derived from the DMP. We distinguished these regions based on the two key properties of the DMP. The high-performing regions for achieving consensus vary depending on the consensus domain (opinion or spatial) used to evaluate the system. By adjusting the DMP parameters, agents can adapt their performance and navigate the various tradeoffs we have identified. We elaborated on some of these tradeoffs such as information versus integration, and generalization versus specialization. Generally, extremes in the number of Messengers (either too few or too many) prove sub-optimal. Pushing the system towards them results in either excessive exploitation or random behavior. The high-performing settings were located in the intermediate parameter regions. We have categorized these regions into three distinct groups, each distinguished by the switching frequency of the DMP. 
\par
The range of switching frequency spans from no switching to excessively fast switching. Agents at these extremes are identified as being specialists or generalists, respectively. Compared to the Messengers ratio, the dimension of switching frequency introduces a more nuanced trade-off. We found that even a minimal number of Messengers can reestablish consensus in collectives of specialists, i.e., when agents maintain fixed states throughout the simulation. On the other hand, faster switching can result in two distinct high-performing behaviors: improved exploration for lower Messenger ratios; or conservation of information at higher Messenger ratios, resulting in a slow update of opinions. 
In the former, the fast-switching Messengers perform random walks long enough to help the collective escape the local optima, corresponding to echo chambers. The latter is the case for collectives mainly composed of Messengers, who update their opinion quite rarely and integrate information only for a short duration. However, this duration is sufficient for them to process information and reach a consensus.
\par
We also highlighted the transient and temporal aspects of collective behavior caused by the DMP. This perspective gives insight into understanding and studying the robustness of the performance against non-stationary behavior or dynamic environments, particularly in scenarios where the initial conditions of the system are different than the expected stationary properties. For example, when the system starts with no Messengers and gradually increases the Messenger population to reach its expected value. Such analyses provide insights for the design of corresponding systems and the selection of parameters that are more robust against transient behaviors. Moreover, we simulated the model in various information distributions with both uni- and multi-modal shapes to assess the dependency of behavior on specific cases. The results showed negligible qualitative differences across different information distributions, demonstrating the generality of our results. 
\par
\newerChanges{
To support our findings, we also conducted two analytical studies under simplifying assumptions and without the effect of Messengers. First, in the discrete case and disregarding the spatial configuration of the system, we analyzed the fixed-point solution of the opinion dynamics. The analysis confirms that a connected interaction network guarantees global convergence, whereas limited connectivity leads to local consensus within disconnected components. This fragmentation highlights the emergence of multiple opinion clusters as a direct consequence of network sparsity.
}
\newerChanges{
Second, we investigated a continuous limit of the spatial opinion dynamics to gain further insight into the conditions under which spatial clustering arises. The analysis shows that under short communication ranges and low diffusion, random local fluctuations can grow and give rise to stable spatial clusters, i.e., echo chambers. These echo chambers emerge even in homogeneous landscapes and are further amplified in heterogeneous environments, where the attractor landscape reinforces the separation of opinions. Messenger agents do not experience homophilic drift and instead move freely while retaining their opinions. This mobility allows them to transmit information across spatial regions, effectively increasing the diffusion of opinion. By introducing dissonance within otherwise isolated clusters, Messengers weaken the positive feedback that sustains fragmentation. This process promotes the merging of local echo chambers and helps the collective converge toward a global estimate, even in heterogeneous environments.
}
\par
While only varying the DMP parameters and keeping the other parameters of the model constant has allowed for an in-depth exploration of the system, future research could expand on this by investigating the influence of the other parameters of the model. This includes aspects such as collective density, arena size, or information noise. Also, investigating the one-dimensional version, which resembles classical opinion dynamics models, could provide valuable insights into a range of related issues in the field. 
\newerChanges{Another promising extension would be to introduce heterogeneous communication ranges, enabling the emergence of centrality and directionality in the network, which are commonly studied in network-based opinion dynamics~\cite{mengers2024leveraging} but not yet explored in our spatial framework.}
An important addition to the analysis of these systems is under dynamic information landscapes, where the collective needs to rearrange itself and adapt to the changing environment. We expect that in such settings, the temporal properties of the system would be highlighted even more.
\par
\newerChanges{In summary, this work introduced a spatially grounded model of collective opinion dynamics that reveals how local homophily and conformity mechanisms, when operating in sparse interaction regimes, can lead to the emergence of echo chambers. At the same time, the spatial embedding enabled a new form of emergent spatial behavior in which agents self-organize along patterns of the information landscape, a process we refer to as contour capturing. To break echo chambers, we proposed the use of heterogeneous agent roles, where a subset of agents, called Messengers, serve as mobile information carriers without updating their own opinions. While our model is abstract and focuses on spatially constrained interactions and mobility, rather than the long-range, algorithmically mediated dynamics typical of online networks~\cite{perra2019modelling}, it offers complementary insights into how opinion dynamics unfold when shaped by limited local proximity and environmental exposure. A~key challenge lies in enabling such role differentiation without centralized control. To this end, we introduced a decentralized switching mechanism based on the Dichotomous Markov Process (DMP). We demonstrated that varying the DMP parameters, and thus the dynamics of role switching, gives rise to rich system-level behaviors and revealed trade-offs, such as specialization versus generalization. These findings suggest that controlled heterogeneity in agent behavior provides a minimal yet powerful strategy for enhancing consensus in spatially embedded collective systems.}
%
%
\section*{Data Availability}
The datasets generated and analyzed during the current study are available in this repository~\cite{raoufi2024simulation}:\\ \url{https://doi.org/10.14279/depositonce-20996}. 
\section*{Code Availability}
The underlying code for this study is available in the GitHub repository and can be accessed via this link:\\ \url{https://github.com/mohsen-raoufi/messengers}. 
\section*{Acknowledgements}
This study was funded by the Deutsche Forschungsgemeinschaft (DFG, German Research Foundation) under Germany’s Excellence Strategy – EXC 2002/1 “Science of Intelligence” – project number 390523135. The funder played no role in the study design, data collection, analysis and interpretation of data, or the writing of this manuscript. We used OpenAI’s ChatGPT to assist with language refinement, grammar correction, and editing the manuscript. All scientific content, analysis, and interpretation were solely developed by the authors.
\section*{Author Contributions}
M.R. contributed to conceiving and implementing models and experiments, as well as to analyzing the results. H.H. and P.R. equally supervised the project. M.R. wrote the first draft of the manuscript with input from all authors. All authors reviewed and revised the manuscript.
\section*{Competing Interests}
All authors declare no financial or non-financial competing interests. 

\bibliography{references}

\end{document}